# Electrically tunable polarizer based on two-dimensional orthorhombic ferrovalley materials

Xin-Wei Shen[1,†], Wen-Yi Tong[1,†], Shi-Jing Gong[1], Chun-Gang Duan[1,2,*]

[1] *Key Laboratory of Polar Materials and Devices, Ministry of Education, East China Normal University, Shanghai, Shanghai 200241, China*

[2] *Collaborative Innovation Center of Extreme Optics, Shanxi University, Taiyuan, Shanxi 030006, China*

(Date: 20 June 2017)

**E-mail:** cgduan@clpm.ecnu.edu.cn

[†]: Cofirst authors

**Abstract**

The concept of ferrovalley materials has been proposed very recently. The existence of spontaneous valley polarization, resulting from ferromagnetism, in such hexagonal two-dimensional materials makes nonvolatile valleytronic applications realizable. Here, we introduce a new member of ferrovalley family with orthorhombic lattice, i.e. monolayer group-IV monochalcogenides (GIVMs), in which the intrinsic valley polarization originates from ferroelectricity, instead of ferromagnetism. Combining the group theory analysis and first-principles calculations, we demonstrate that, different from the valley-selective circular dichroism in hexagonal lattice, linearly polarized optical selectivity for valleys exists in the new type of ferrovalley materials. On account of the distinctive property, a prototype of electrically tunable polarizer is realized. In the ferrovalley-based polarizer, a laser beam can be optionally polarized in *x*- or *y*-direction, depending on the ferrovalley state controlled by external electric fields. Such a device can be further optimized to emit circularly polarized radiation with specific chirality and to realize the tunability for operating wavelength. Therefore, we show that two-dimensional orthorhombic ferrovalley materials are the promising candidates to provide an advantageous platform to realize the polarizer driven by electric means, which is of great importance in extending the practical applications of valleytronics.

**Keywords:** valleytronics, 2D materials, ferroelectricity, ferrovalley material

# 1. Introduction

Since the discovery of graphene [1], the interests in exploring novel two-dimensional (2D) materials rooted in hexagonal lattice have been intensively stimulated [2-5]. Similar to charge and spin, valley index [6, 7], as an emerging degree of freedom of electrons, constitutes the binary states in solids, leading to valleytronic devices on information functional applications [8-11]. Very recently, Tong *et.al* [12] introduced the concept of ferrovalley material in transition metal dichalcogenides. Compared with valley degeneracy splitting through external means [13-17], the valley polarization in the proposed ferrovalley monolayers, originating from ferromagnetism, is spontaneous. It is, thus, of great importance in paving the way to the practical use of valleytronics in a nonvolatile way. Interestingly, due to the polarized optical selectivity for valleys, chirality-dependent optical band gap exists here, which indicates the possibility to judge the valley polarization utilizing noncontact and nondestructive optical ways, and offers the additional opportunities for ferrovalley materials as optical devices.

However, the valley polarization in previous study [12] originates from the inherent exchange interaction, making an external magnetic field necessary for the hexagonal ferrovalley monolayer to control its polarity. Compared with the control of degrees of freedom by the energy-intensive magnetic way, manipulation via purely electric fields is highly desirable due to its advantages of ultra-high speed and ultra-low power consumption. In this context, if we could connect ferrovalley with ferroelectricity, such an additional mechanism for intrinsic valley polarization would provide the opportunity to realize the attractive electrical control of ferrovalley states.

As we know, valley-related research mainly focused on pioneering 2D materials with hexagonal lattice [18-20]. However, multi-valley band structures are proved to exist in other crystallographic systems, such as the orthorhombic phases [21]. They provide a promising and entirely new platform for the studies of the fundamental physics in valleytronics, and more importantly, make the realization of tuning valley degree of freedom through electric means possible. It is then naturally to raise the following questions: Could the robust ferrovalley states persist in new kinds of crystallographic systems? How to dynamically utilize the valley polarization on the modulation of ferroelectricity? Would the new member of valleytronic materials demonstrate distinctive valley-related optical properties, other than the valley-selective circular dichroism in hexagonal systems [11, 12, 18]?

In this Letter, we propose the monolayer group-IV monochalcogenides in ferroelectric state as an entirely new ferrovalley member, extending the concept of ferrovalley materials from hexagonal to orthorhombic systems. Here, the spontaneous valley polarization is induced by ferroelectricity, instead of ferromagnetism. Taking the monolayer GeSe as an example, our analytical research and first-principles calculations show that in this system the optical transitions at valleys are linear polarization dependent. Consequently, linear polarization-dependent optical band gap could be observed in these ferrovalley materials. The interesting non-degenerate optical excitation offers the possibility to realize a prototype of electrically tunable polarizer. In the polarizer, *x*- or *y*-polarized light can be optionally emitted, depending on the specific ferroelectric state manipulated by electric fields. Compared with traditional mechanically driven ones, the ferrovalley-based polarizer provides a more accurate and rapid way to electrically control the polarization state of linearly polarized radiation. Additionally, the proposed device based on 2D materials meets the requirement of next-generation electronic products towards miniaturization and multi-functionality. On account of the inverse piezoelectric effect [22, 23], continuous tunability of band gap in monolayer GeSe can be realized via external electric fields. Thus, the more advanced polarizer with advantage of wide operating wavelength range is emerging to further widen the valleytronic applications.

## 2. Methods

Our first-principles calculations are carried out by density-functional theory (DFT) using the projector augmented wave (PAW) [24] method which implemented in the Vienna *ab initio* Simulation Package (VASP) [25, 26]. The exchange correlation potential is described by generalized gradient approximation (GGA) of Perdew-Burke-Ernzerhof (PBE) [27] functions. The cutoff energy of 500 eV is applied to the plane wave basis set. All the structures are optimized until the forces tolerance below 1 meV/Å and self-consistent convergence for electronic energy is $10^{-6}$ eV. For the optical property calculations, we take the own-developed code OPTICPACK [28-32]. The Brillouin zone (BZ) is sampled with a Γ-centered 12×12×1 grid and increase to 24×24×1 for optical calculations. A vacuum space of 15 Å is used to avoid interactions between periodically repeated layers.

## 3. Results and discussion

In analogy to black phosphorus [33], group-IV monochalcogenides [34-41] exhibit waved structures. Figures 1(a) and (b) display the primitive cell of pristine monolayer GeSe with inversion symmetry. There is no relative displacement between cations and anions along $x$ or $y$ axis, indicating an in-plane paraelectric phase, labelled as $P_0$.

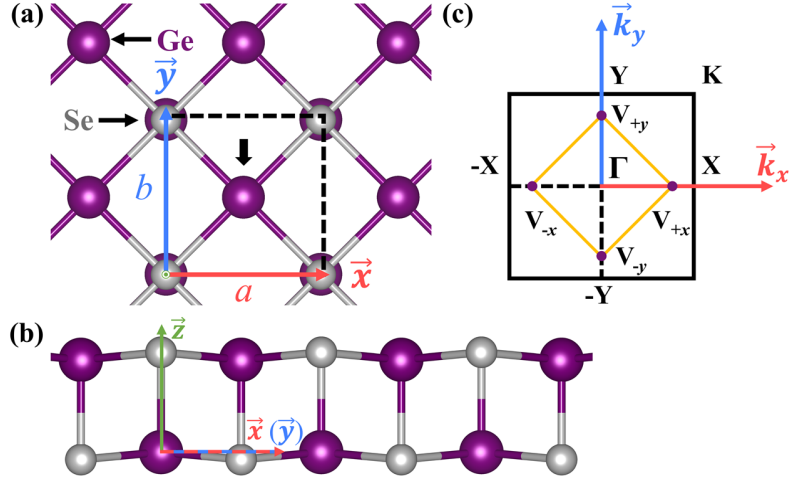

Figure 1. (a) Top and (b) side view of the lattice structure with optimized lattice constants $a = b = 4.05$ Å of paraelectric monolayer GeSe. (c) shows the Brillouin zone. Two pairs of valleys $V_{\pm x}$ and $V_{\pm y}$ locate in the $\pm X$–$\Gamma$ and $\pm Y$–$\Gamma$ symmetry lines, respectively.

As plotted in figure 2(a), the band structure of the nonpolar GeSe obviously displays multi-valley characteristics, where two pairs of valleys locate in the X–$\Gamma$ and Y–$\Gamma$ high-symmetry paths instead of the symmetry corners of the BZ, as illustrated in figure 1(c). They are energetically degenerate with the $C_{2v}$ point group symmetry. Note that valleys $V_{+x(y)}$ and $V_{-x(y)}$ are related by time-reversal operation. Since the spin degree of freedom of electrons is inessential to explore valley physics of the monolayer group-IV monochalcogenides, we just discuss the electronic and optical properties at valleys $V_{+x}$ and $V_{+y}$ as representative. For the $V_x$ ($V_y$) valley, the valance band maximum (VBM) and the conduction band minimum (CBM) are purely occupied by $p_x$ ($p_y$) orbitals of Se and Ge atoms, respectively. In spite of the distinct occupied states, the band gaps of $V_x$ and $V_y$ valleys are identical (about 0.83 eV). On account of the degeneracy between valleys, the paraelectric GeSe monolayer demonstrates the paravalley state, i.e., state without spontaneous valley polarization.

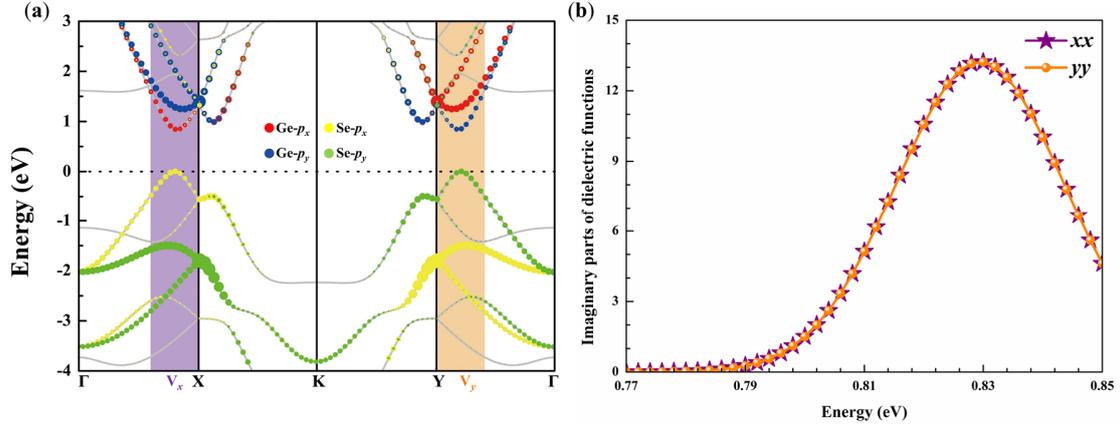

Figure 2. (a) Band structure and (b) the imaginary parts of complex dielectric function $\varepsilon_2$ of paraelectric GeSe monolayer. The radius of solid color circles is proportional to the contributions of specific orbitals: the red and blue one represent $p_x$ and $p_y$ of Ge atoms and the yellow and green one correspond to $p_x$ and $p_y$ of Se atoms. The Fermi level is set to zero. Cases excited by both linearly $x$- and $y$-polarized light are presented in (b).

For paravalley materials with hexagonal lattice, the circularly polarized optical selectivity for valleys has been proposed which is essentially rooted in conservation of overall azimuthal quantum number [42]. A similar approach is adopted here to explore the valley-dependent optical selection rules in the monolayer GeSe. Wavefunctions are constructed as Bloch functions in terms of atomic orbitals $\phi(r-R_j)$ localized at the lattice site $R_j$:

$$\psi = \frac{1}{\sqrt{N}} \sum_j e^{ik \cdot R_j} \phi(r-R_j). \tag{1}$$

Here $N$ is the number of unit cells for the crystal and $k$ is the wavevector. The local atomic states bear different principal quantum number $n$, azimuthal quantum number $l$ and magnetic quantum number $m_l$, then the local wavefunctions can be expressed as:

$$\phi_{n,l,m_l}(r-R_j) = R_{n,l}(r-R_j) Y_{l,m_l}(\theta,\varphi), \tag{2}$$

where $(r, \theta, \varphi)$ are the standard spherical coordinates. In the case of monolayer GeSe, the basis functions of conduction and valance band are $|\psi_c\rangle = |p_x\rangle(|p_y\rangle)$ on Ge and $|\psi_v\rangle = |p_x\rangle(|p_y\rangle)$ on Se at the $V_x$ ($V_y$) valley, accordingly. When the twofold rotation ($C_2$) is applied, the Bloch phase shifts from one lattice site to another, acting on the phase factor $e^{i\mathbf{k}\cdot\mathbf{R}_j}$. In addition, due to the symmetry operation, the rotation of local atomic coordinates of the spherical harmonics gives rise to the desynchronization of the azimuthal phase simultaneously. Taken the two distinct contributions into account, the transformation of electronic states associated with the twofold rotation can be easily deduced. For the adapted symmetry $C_{2x}$ at $V_x$ valley, after tedious yet straightforward calculations, we have

$$\hat{C}_{2x}|\psi_i(\mathbf{V}_x)\rangle = |\psi_i(\mathbf{V}_x)\rangle, i = c, v. \qquad (3)$$

Here, the symbol $c$ and $v$ refer to the CBM and VBM respectively. We notice that basis functions with linear combination of atomic orbitals are invariant through rotational symmetry. Then, the overall azimuthal quantum number $M_x$ is eventually calculable,

$$e^{iM_x(\varphi+\pi)} = e^{iM_x\varphi}. \qquad (4)$$

As a consequence, both the VBM and CBM hold the identical $M_x = 0$. For the optical transition at $V_x$ valley, the angular momentum selection rule indicates that $\Delta M_x = 0$, corresponding to the absorption of $x$-polarized photons with the symmetry of $\cos(\theta)$ in the case. Similarly, under the twofold rotation $C_{2y}$ at $V_y$ valley, we can deduce that $\Delta M_y = 0$, indicating the optical absorption at $V_y$ valley can only be excited by the $y$-polarized light. Therefore, based on conservation of overall angular momentum, the valley-dependent optical selection rule in paraelectric monolayer GeSe could be summarized as the following: at $V_x$ valley the optical transition from VBM to CBM could only be excited by $x$-polarized light, while in $V_y$ valley it couples to the radiation polarized along $y$-direction. It is interesting to point out that the valley-dependent optical selection rule in such orthorhombic material is distinct from that of the hexagonal systems. It corresponds to linearly polarized light here rather than the circularly one in graphene and transition metal dichalcogenides.

Because of the specific linearly optical selection rule here, as shown in figure 2(b), the calculated optical curves related to $x$- and $y$-polarized light are completely overlapped. Their identical optical band gap with the magnitude of 0.83 eV is in accordance with the electronic one gained from figure 2(a). We note that the degeneracy of optical properties excited by $x$- and $y$-polarized radiation is due to the paravalley state in such monolayer group-IV monochalcogenides.

When the intrinsic ferroelectricity of monolayer GeSe is taken into account, as shown in figures 3, the relative displacement between Ge and Se atoms occurs, making its inherent inversion symmetry broken. The ferroelectric states with polarization direction along *x* and *y* axis are labbelled as $P_x$ (figure 3(a)) and $P_y$ (figure 3(b)), accordingly. The spontaneous in-plane polarization ($P_s$) estimated in monolayer GeSe can reach to $3.5 \times 10^{-10}$ C/m, which is well consistent with previous works [39-41].

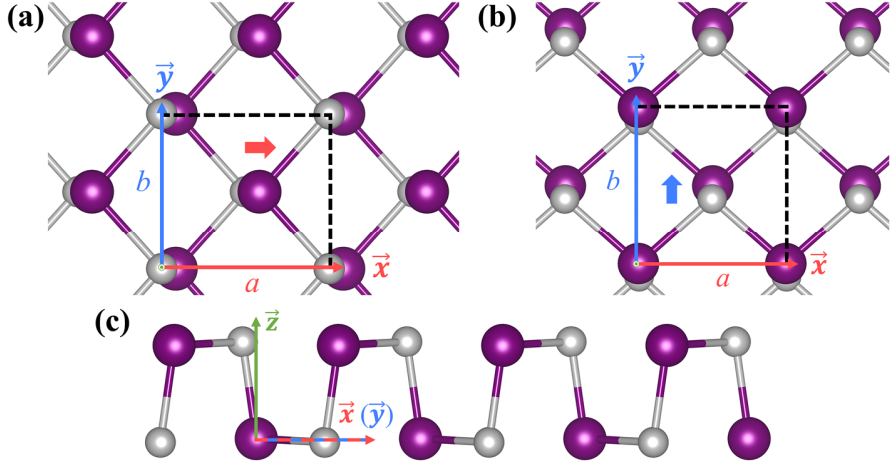

Figure 3. Top view of the lattice structure for monolayer GeSe with ferroelectric polarization along (a) *x* axis and (b) *y* axis. The arrows point out the orientation of polarization. For the ferroelectric phase $P_x$ ($P_y$), the primitive cell is inequiaxial with optimized lattice constant *a* (*b*) = 4.27 Å and *b* (*a*) = 3.98 Å. (c) presents the side view of monolayer GeSe in ferroelectric state.

Their band strucutres are plotted in figures 4(a) and (b). Compared with electronic distribution in paraelectric GeSe monolayer, the occupied states of VBM and CBM at valleys $V_x$ and $V_y$ are nearly unchanged. Nevertheless, degeneracy between two pairs of valleys is lifted. Along with the spatial inversion asymmetry, the band gaps of $V_x$ and $V_y$ are no longer identical, which is the critical sign of spontaneous valley polarization. Therefore, the intrinsic ferrovalley with nonvolatile polarization state has been induced by ferroelectricity in GeSe monolayer. As clearly shown in figure 4(a), for the ferroelectric GeSe monolayer under $P_x$ case, the global band gap locates in $V_x$ valley with the magnitude of 1.14 eV. While, $V_y$ valley owns a larger band gap ~1.67 eV. Interestingly, although the $V_x$ valley still belongs to $C_{2v}$ point group with the *x*-polarized light selective excitation, ferroelectric polarization along *x* direction makes the symmetry of $V_y$ valley reduced. Now, it holds the $C_s$ point group, where the optical transition from VBM to CBM can be excited by both the *x*- and *y*-polarized radiation. As a result, the occurrence of energetically non-degenerate valleys implies decoupled optical band gaps of linearly polarized light.

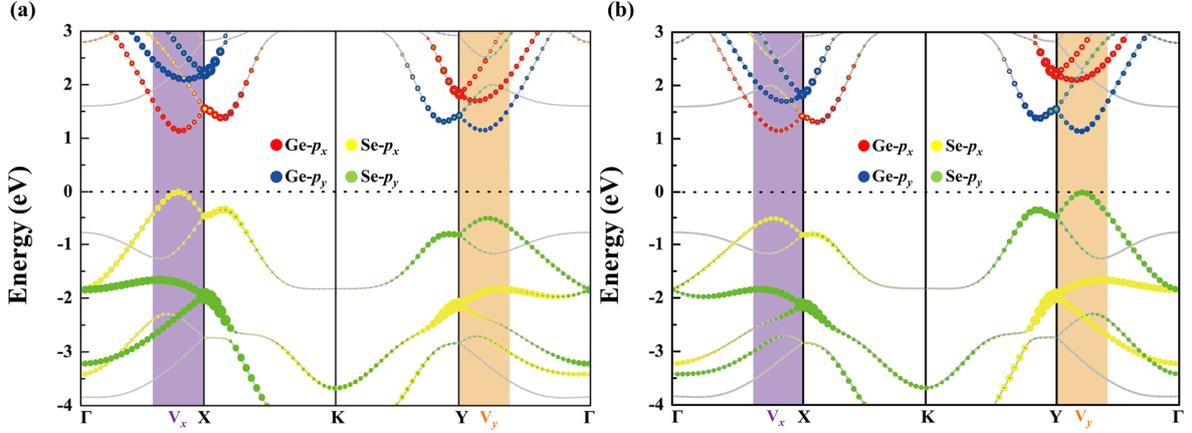

Figure 4. The band structure of ferroelectric phase GeSe monolayer in (a) $P_x$ and (b) $P_y$ state. The radius of solid color circles is proportional to the contributions of specific orbitals: the red and blue one represent $p_x$ and $p_y$ of Ge atoms and the yellow and green one correspond to $p_x$ and $p_y$ of Se atoms. The Fermi level is set to zero.

Figure 5(a) shows the optical properties in the $P_x$ state. As expected, the optical band gap related to $y$-polarized light experiences a blue shift compared with that of the $x$-polarized one. The optical absorption starts from the energy of 1.14 eV, corresponding to the direct optical band gap at $V_x$ valley.

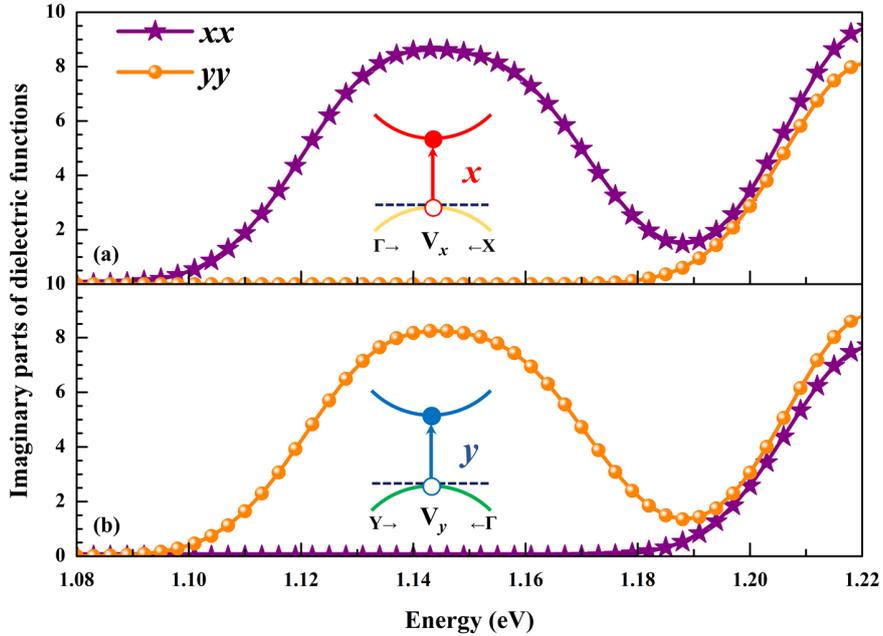

Figure 5. The imaginary parts of complex dielectric function $\varepsilon_2$ excited by linearly $x$-polarized light and $y$-polarized light of ferrovalley GeSe monolayer in (a) $P_x$ and (b) $P_y$ state.

When the ferroelectric polarization occurs along $y$ direction, the polarity of valley polarization is reversed in the meantime. The global band gap with the same magnitude as $P_x$ state now shifts to $V_y$ valley (see figure 4(b)). Group theory analysis indicates that, in such case, the $V_y$ valley reverts to its original symmetry with $C_{2v}$ point group, whose interband transition between VBM to CBM can be merely excited by the $y$-polarized light. For the $V_x$ valley with lower symmetry, $x$- and $y$-polarized radiation belong to the identical irreducible representation, making the optical transition couple to the double-degenerate in-plane linearly polarized radiations. The energetical non-degeneracy between valleys is obviously verified as unequal optical band gaps between $xx$ and $yy$ components in figure 5(b).

In a word, the intrinsic ferroelectricity in monolayer GeSe breaks the spatial symmetry and lifts the valley degeneracy. The valley polarization, then, naturally appears in the ferrovalley state, whose polarity is optionally controlled by the direction of the ferroelectric polarization. The ferrovalley states existing in such orthorhombic systems break through the restriction of conventional hexagonal lattice. The valley-related properties in the new member of ferrovalley family seem peculiar and quite different from previous ones based on hexagonal lattice. Table I lists their comparisons.

Table 1. Comparisons of two kinds of valleytronic materials with associated physical phenomena. As a representative of ferrovalley, monolayer $VSe_2$ and GeSe are listed below to make a contrast.

|  | $VSe_2$ | GeSe |
| --- | --- | --- |
| Mechanism for valley polarization | Ferromagnetism | Ferroelectricity |
| Lattice structure | Hexagonal | Orthorhombic |
| Optical properties | Circular dichroism | Linear dichroism |
| Anomalous valley Hall effect | √ | × |

For monolayer $VSe_2$ [12], the spontaneous valley polarization is induced by ferromagnetism. Here, in monolayer GeSe, it couples to the primary ferroelectricity, indicating an entirely new mechanism for ferrovalley physics. What's more, linearly polarized optical selectivity for valleys in such new ferrovalley materials gives rise to an additional relationship between optical field and valley index, rather than the valley-selective circular dichroism reported in hexagonal monolayers. As we know, Berry curvature drives an anomalous transverse velocity for carries around valley [19, 43], which directly causes the existence of valley Hall effect [6, 10, 18] and the more exciting anomalous valley Hall effect in ferrovalley VSe$_2$ [12]. By calculating the Berry curvatures for monolayer GeSe based on the Kubo formula [44], we find that due to the point group symmetry, majority carriers bear zero Berry curvature and then pass through the Hall bar straightforwardly without transverse deflection. In addition, pairs of valleys protected by time-reversal symmetry, i.e. valleys V$_{+x(y)}$ and V$_{-x(y)}$, are energetically degenerated, which restricts the possibility of transverse charge Hall current. We note that the anomalous valley Hall effect is absent in both slightly n- and p-doped ferrovalley materials with orthorhombic lattice.

In consideration of linearly polarized optical selectivity for valleys, optical approach is an effective method with advantage of noncontact and nondestructiveness to determine the occurrence of valley polarization and its polarity reversal in such ferrovalley monolayers. In turn, the specific ferrovalley state, which depends on the orientation of electric dipole and is switchable via external electric fields [45], offers the possibility to tune the direction of linearly polarized light. An idea for the unique valleytronic device application is thus inspired. Figure 6 displays the prototype of electrically controllable polarizer based on the monolayer GeSe.

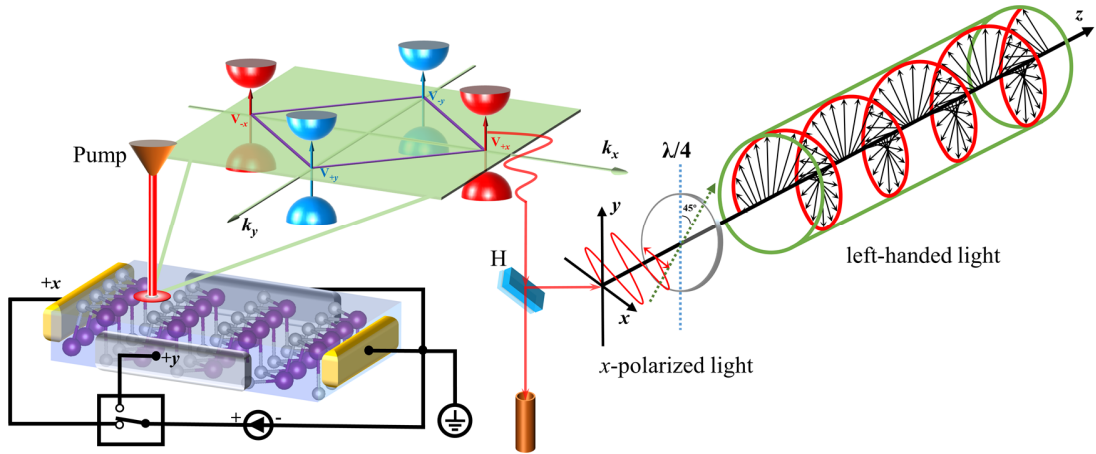

Figure 6. Proposed electrically tunable polarizer based on the ferrovalley GeSe monolayer. In the polarizer, $x$- or $y$-polarized light can be optionally generated, depending on the ferrovalley state manipulated by external electric fields. By additionally placing a quarter-wave plate, specific chirality of circularly-polarized light is available and reversible via electric means. H is a half mirror to transmit the polarized light for different demands.

In the polarizer, a laser beam with the energy of 1.14 eV is incident on the pristine monolayer GeSe.

The valley polarization of monolayer GeSe could be induced by an external electric fields. When the in-plane field is applied along the $x$ axis, the band gap locates in the $V_x$ valley, where electrons jump to the excited state by absorbing the laser beam. In consideration of the photoluminescence process, the excited electrons in the $V_x$ valley finally re-emit radiations as it drops back to the ground state. Due to the optical selectivity for the $V_x$ valley, the excitation light is polarized in $x$ direction. If the double-throw switch applies the electric field along $y$ axis, monolayer GeSe undergoes a ferroelectric phase transition from $P_x$ to $P_y$ state. Polarity of its valley polarization simultaneously reverses. Now, the $V_y$ valley possesses the smallest band gap. Its symmetry indicates the emission of $y$-polarized radiation. Hence, through controlling the direction of external electric fields, ferroelectric phase, and then valley polarization of monolayer GeSe could be manipulated. Generally, the polarization switching in a ferroelectric behaves an ultrafast dynamics, of which the time scale can vary from microseconds to nanoseconds [46-48]. With the linearly polarized optical selectivity, the ferrovalley-based polarizer offers the possibility to dynamically polarize the laser beam in $x$- or $y$-direction in a more accurate and high speed electrical way.

Furthermore, if a set of quarter-wave plate is placed perpendicular to the propagation of linearly polarized light with its optical axis deviating 45° anti-clockwise from the $x$ direction (or equivalently clockwise from the $y$ direction), then chirality of circularly polarized light could be electrically tunable in a similar way. The linearly polarized light in $x$ direction, which is emitted by applying the electric field along $x$ axis, transfers to the left-handed radiation. Converting the switch makes the monolayer GeSe in $P_y$ state. Then, the generated circularly polarized light consequently becomes the right-handed one.

Previous analysis implies that utilizing the ferroelectricity in monolayer group-IV monochalcogenides, a laser beam with specific wavelength can be electrically polarized to linearly and even circularly polarized radiation. The polarization state of the light is optionally tuned via advantageous electric means. More interestingly, the giant inverse piezoelectric effects and strain-sensitive band gap [49-52] in these materials indicate the potential to realize an advanced polarizer with wide operating wavelength range.

To further study the influence of epitaxial strain on the band gap of ferrovalley GeSe, we change the lattice constant $a$ ($b$) in the ferroelectric polarization for $P_x$ ($P_y$) state within the range 4.18 – 4.35 Å, which is equivalent to the strain from -2.0% to +2.0%. The lattice constant $b$ ($a$), as well as atomic coordinates, are optimized for each case. As shown in figure 7, when the internal generation of a mechanical strain, resulting from an applied electric field, varies from the compressive to the tensile one, the band gap of monolayer GeSe increases linearly.

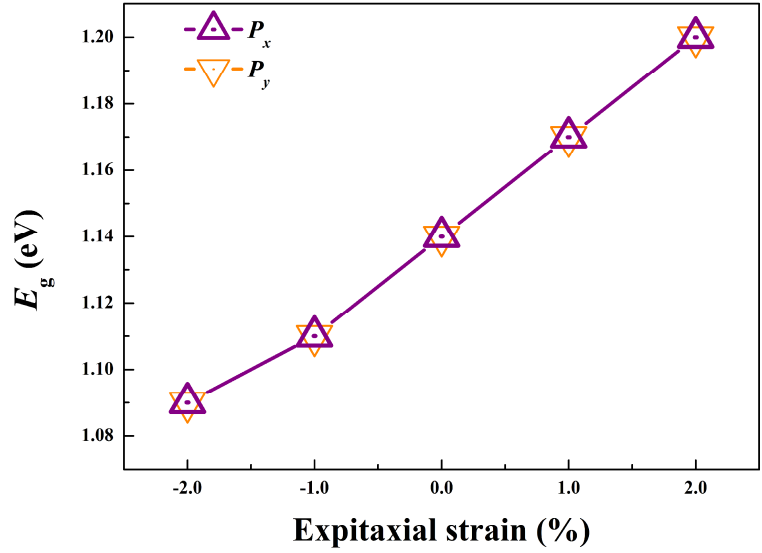

Figure 7. The band gap of monolayer GeSe in ferrovalley state with respect to epitaxial strain from -2.0% to +2.0%.

For the $P_x$ state, strain makes few effect on location of the direct band gap in $k$-space. It is always at $V_x$ valley, corresponding to $x$-polarized radiation. Whereas, the value of the band gap greatly changes, from the maximum of 1.20 eV under tensile 2% strain to the lowest ~1.09 eV under the same magnitude of compressive one, which equals to the enhancement of 105 nm in wavelength for the emission. Note that the band gap of monolayer GeSe in $P_y$ state with respect to epitaxial strain is identical to the case in $P_x$ state. However, it locates in the $V_y$ valley with the $y$-polarized light selective excitation. As a result, electric-field-induced strain engineering demonstrates an effective way to change the direct or equivalently optical band gap of monolayer GeSe in ferrovalley state. The wavelength for emergent light could be controlled in a quite large range when the mechanical strain varies within a small scale. Such effect implies that through the external electric fields, we can not only generate either linearly polarized light or circularly polarized light as demands, but also acquire the polarized lights with various wavelengths.

## 4. Conclusion

In conclusion, we introduce a new family member of ferrovalley materials in orthorhombic lattice, where spontaneous valley polarization can be induced by intrinsic in-plane ferroelectricity. In consideration of linearly polarized optical selectivity for valleys in such monolayer group-IV monochalcogenides, we propose an electrically tunable polarizer, by which linearly polarized light or even circularly polarized one can be acquired and optionally tuned through advantageous electric means. Furthermore, band gap in these systems is sensitive to epitaxial strain, indicating the possibility to continuously control the operating wavelength of the polarizer in a large scale. The new kind of 2D materials with spontaneous valley polarization widens the members of ferrovalley family, and is of great potentialities in valleytronic and optoelectronic applications.

## 5. Acknowledgments

This work was supported by the National Key Project for Basic Research of China (Grant Nos. 2014CB921104 and 2013CB922301), the National Natural Science Foundation of China (Grant No. 51572085). Computations were performed at the ECNU computing center.